\documentclass[aps,superscriptaddress,nofootinbib]{revtex4}
\usepackage{amsmath}
\usepackage{graphicx}
\usepackage{caption}
\usepackage{color}
\usepackage{slashed}

\begin{document}

\title{Thoughts on entropic gravity in the Parikh-Wilczek tunneling model of Hawking radiation}

\author{Wen-Yu Wen}\thanks{%
E-mail: steve.wen@gmail.com}

\affiliation{Department of Physics and Center for High Energy Physics, Chung Yuan Christian University, Chung Li City, Taiwan}
\affiliation{Leung Center for Cosmology and Particle Astrophysics\\
National Taiwan University, Taipei 106, Taiwan}

\begin{abstract}
In this letter, we use the Parikh-Wilczek tunneling model of Hawking radiation to illustrate that a reformulation of Verlinde's entropic gravity is needed to derive the Newton's law for a temperature-varying screen, demanded by the conservation of energy.  Furthermore, the entropy stored in the holographic screen is shown to be additive and its temperature dependence can be obtained.
\end{abstract}

\maketitle



\section*{Verlinde's entropic gravity and holographic screen}
To explain why gravity is so different from other three forces in the nature, Verlinde proposed that gravity could be regarded as an entropic force caused by changes in the information associated with the positions of material bodies\cite{Verlinde:2010hp}.   His idea can be illustrated by a particle with mass $m$ approaching some hypothesized holographic screen.  The screen bounds the emerged part of space containing the particle, and stores data that describe the part of space that has not yet emerged, as well as some part of the emerged space.  The change in entropy on the screen, denoting as $\Delta S$, is assumed to be linear in the particle's displacement $\Delta x$ as follows:
\begin{equation}\label{eqn:entropy_displace}
\Delta S = 2\pi k_B \frac{mc}{\hbar}\Delta x
\end{equation}
If one further assumes that the screen is thermalized at the Unruh temperature set up by particle's acceleration, then the Newton's second law can be {\sl derived} from the entropic force relation:
\begin{equation}\label{eqn:verlinde-relation}
F\Delta x = T \Delta S.
\end{equation}

One opposition to this proposal is given by \cite{Kobakhidze:2010mn} using the measurement result of quantum states of ultra-cold neutron under the Earth's gravity.  According to Kobakhidze's argument, a pure state neutron would have evolved into a mixed state thanks to $\Delta S >0$ in (\ref{eqn:entropy_displace}).  This criticism, however, is questioned in \cite{Chaichian:2011xc} and one resolution was suggested in \cite{Abreu:2013rxe} by abandoning an implicit assumption in \cite{Kobakhidze:2010mn} that the entropy on holographic screen is additive.  Instead, unitarity could still be restored even with $\Delta S>0$ if bits on the screen were entangled in some delicate way.  Nevertheless, it remains unclear how the entropy is entangled on the holographic screen.  

The other restriction of (\ref{eqn:verlinde-relation}) is that a holographic screen at thermal equilibrium corresponds to the particle moves in uniform acceleration.  A generalization of entropic force relation is needed to describe a particle with generic motion.   Similar to a generalization of Newton's second law to include varying mass, it is natural to generalize the entropic gravity to include a varying screen temperature.  With that being said, we consider an adiabatic process such that {\sl temperature} of holographic screen is slowly varying but still well defined while a massive particle moves relatively to the screen.  The generalized entropic force formula then becomes
\begin{equation}\label{eqn:work-thermal-relation}
F\Delta r = \Delta (TS),
\end{equation}
for varying screen temperature $T$ and screen entropy $S$ while a particle is displaced a distance $\Delta r$ from the screen at fixed location.

In the following, we will illustrate that Verlinde's screen could be realized in the tunneling model of Hawking radiation if a reformulation of entropic gravity is adopted to compensate the effect of varying temperature.  Following a similar argument as in \cite{Kobakhidze:2010mn} but for temperature-varying screen, we show that the entropy can still be additive on the screen and its temperature dependence can be derived. 

\section*{Entropic gravity with temperature-varying screen}
The original treatment of Hawking radiation by Hawking is to consider perturbation in a fixed background of Schwarzschild black hole.  The thermal spectrum brought up controversial debates over the Information Loss Paradox.  Parikh and Wilczek considered radiation as an outgoing tunneling particle where the conservation of energy is enforced\cite{Parikh:1999mf}.  It was later confirmed that in their model information is also conserved by computing the mutual information for two successive radiations\cite{Zhang:2009jn}.   If gravity has an entropic origin, it is desirable to describe the Hawking radiation (as a tunneling process), at least for Schwarzschild black holes, as some form of entropy change in a holographic screen.  We recall that back reaction from radiated particle to the black hole makes it a system away from thermal equilibrium, namely temperature varying or not well defined.  As a result it is proper to apply the general formula (\ref{eqn:work-thermal-relation}) rather than  (\ref{eqn:verlinde-relation}) to the tunneling model.  For a screen which stores same amount of information as the black hole entropy, and possesses temperature same as the black hole, the change on right-hand-side before and after radiation can be computed as 
\begin{eqnarray}\label{thermal_mass}
\Delta (TS) = \frac{1}{8\pi G (M-\omega)}\cdot 4\pi G (M-\omega)^2 \nonumber\\
-\frac{1}{8\pi GM}\cdot 4\pi GM^2 = -\frac{\omega}{2},
\end{eqnarray}
which can be understood as a quantum of mass(energy) $\omega$ that is discarded from the black hole up to a factor $1/2$.\footnote{This factor has been pointed out in the \cite{Padmanabhan:2009kr} as equipartition of energy in the horizon degrees of freedom. In brief, each quanta of area fluctuation contributes an energy $T/2$, that is $E = \frac{1}{2}nT$ for $n=A=4S$.  One can derive (\ref{thermal_mass}) for loss of energy (mass) $\omega$.}  The left-hand-side computes the required work by pulling a point-like varying mass(energy) from the black hole.  According to the Newton's law of gravity, that is 
\begin{equation}
\int_{2GM}^{2G(M-\omega)}F(r) dr = -\frac{G(M-\frac{r}{2G})(\frac{r}{2G})}{r}\big|_{2GM}^{2G(M-\omega)}=-\frac{\omega}{2},
\end{equation}
which is consistent with the right hand side.  Therefore, the gravitational force can again be identified with an entropic force, even for a nonequilibrium system.

\section*{Holographic screen in the Parikh-Wilczek tunneling model of Hawking radiation}
Inspired by the reformulation of entropic gravity in the previous discussion, we are motivated to study of a holographic screen at arbitrary position but admitting varying temperature, in reaction to displacement of a radiation particle.   For symmetry reason, a spherical screen enclosing the black hole seems a good choice.  We denote the information stored on a holographic screen associating to a black hole with mass $M$ as $S_M(r)$.  The screen locates at some distance $r$ from the center of black hole.  We may want to have $r>2GM$ to avoid ambiguity of a screen {\sl inside} the black hole.  Before proceeding with the general formula (\ref{eqn:work-thermal-relation}), we would like to first point out some difficulties appeared in the original formulation of entropic gravity (\ref{eqn:verlinde-relation}) in the tunneling model of black hole radiation. 
If we assume the entropy stored on the screen follows simple coarse grained relation as shown in the Figure $1$, following the neutron-earth system in \cite{Kobakhidze:2010mn}:

\begin{figure}[tbp]

\includegraphics[width=0.6\textwidth]{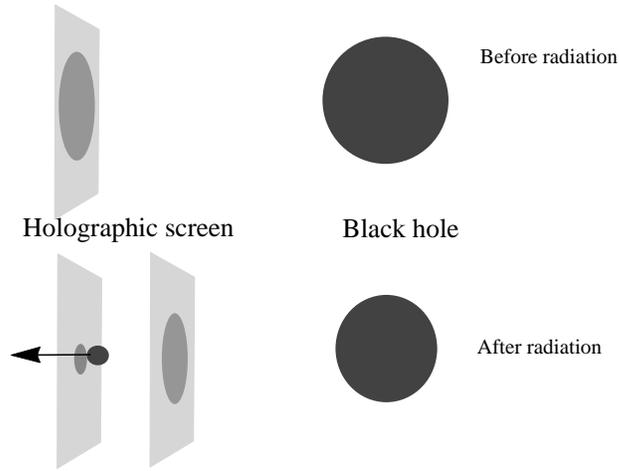} 
\caption{\label{fig:coarse} A schematic drawing of a particle radiating out of a black hole and their holographic screen.}

\end{figure}

\begin{equation}\label{eqn:coarse-grained-relation}
S_M(r+\Delta r) = S_\omega (r+\Delta r) + S_{M-\omega}(r),
\end{equation} 
where $\Delta r$ is the distance between radiated mass and the screen.  For infinitesimal displacement, we have $S_M(r+\Delta r) \simeq S_M(r) + \Delta S$.  If entropy on the screen were additive, that is,
\begin{equation} \label{eqn:additive}
S_{M-\omega}(r) = S_M(r)- S_\omega(r)
\end{equation}
One obtains
\begin{equation}
\Delta S_M \simeq S_{\omega}(r+\Delta r) - S_{\omega}(r).
\end{equation}
Since $\Delta S_M \propto \Delta r$ for its entropic origin, we come to the same conclusion that the translation operator is no longer Hermitian and disobeying quantum mechanics.  In the \cite{Abreu:2013rxe}, a generalized entropy formula by Tsallis\cite{Tsallis:1988} was called to restore the unitarity of translation, where an additional entangled term was essential.  In our radiation-black hole system, it was known that the Parikh-Wilczek model predicted a non-thermal radiation spectrum and  correlation between two emissions $\omega_1,\omega_2$ was computed as $8\pi \omega_1 \omega_2$, known as mutual information\cite{Zhang:2009jn}.  The entanglement is necessary for  the conservation of entropy.   Therefore, it is possible to correct the additivity condition (\ref{eqn:additive})  to include a term $S_{ent}>0$ encoding entanglement between black hole and radiation:
\begin{equation}\label{eqn:nonadditive}
S_{M-\omega}(r) = S_M(r)- S_\omega(r)-S_{ent}(r),
\end{equation}
such that entropy is conserved before and after radiation\footnote{Classically, a black hole is forbidden to split because $S_M(r) > S_{M-\omega}(r) + S_\omega(r)$.  In the Parikh-Wilczek model, the entanglement provides additional source of entropy to conserve the entropy during evaporation process.}.  Notice the relative sign for the last term is different from usual entanglement between two subsystems.  This difference in fact makes situation worse since now $\Delta S_M \simeq S_\omega(r+\Delta r)-S_\omega(r) - S_{ent}(r)$.  For a translation to be unitary, say $S_\omega(r+\Delta r) = S_\omega(r)$, one has $\Delta S_M <0$ implying a violation of second law of thermodynamics, which reflects the fact that black hole cannot evaporate classically.  If  the Hawking radiation were thermal, one might expect the increased entropy during thermal process is large enough to compensate the reduced entropy due to splitting of radiated particle from black hole.  However, in the Parikh-Wilczek's tunneling model, one has no such additional source for entropy since the process is nonthermal.

Now we turn to application of the general entropic formula (\ref{eqn:work-thermal-relation}).  Because temperature may vary for screens at different locations, we assume the coarse grained relation (\ref{eqn:coarse-grained-relation}) is modified as 
\begin{equation}
T(r+\Delta r) S_M(r+\Delta r) = T(r+\Delta r)S_\omega(r+\Delta r) + T(r)S_{M-\omega}(r).
\end{equation}
If we assumes a linear relation between screen temperature and small displacement, that is $T(r+\Delta r) \simeq T(r) + \Delta T$, one obtains 
\begin{equation}
\Delta S_M \simeq  [S_\omega(r+\Delta r)-S_\omega(r)]+[\frac{\Delta T}{T}S_\omega(r+\Delta r)-S_{ent}.]
\end{equation}
For translation of the radiated particle being unitary, $S_\omega(r)=S_\omega$ is a constant.  Vanishing of $\Delta S_M$ for $\Delta r=0$ suggests $S_{ent}(r)=0$.  According to (\ref{eqn:nonadditive}), the entropy on screen is additive.   The temperature dependence can also be obtained by integration:
\begin{equation}\label{eqn:screen-temperature}
S_M(r)-S_M(r_0) = S_\omega \ln\frac{T(r)}{T(r_0)},
\end{equation}
 for some reference position $r_0$.  The movement of radiated particle with respect to the screen does not change corresponding entropy $S_\omega$. Instead, it feels different temperatures from screens at different locations.  If one recalls the scenario of a massive particle tunneling out of black hole horizon: the radiated particle experiences less gravitational pull while it moves away from the black hole, as if it feels colder from holographic screen.  The screen temperature reflects the acceleration or deceleration as what happens in the Unruh effect.

\section*{Discussion}
In this letter, with an explicit example of Parikh-Wilczek's tunneling model of Hawking radiation, we illustrate that Verlinde's holographic screen could have violated unitarity of quantum mechanics if it were additive in entropy.  This sickness cannot be simply cured by assumption that the entropy on screen is entangled.  Motivated by a general formula of entropic gravity, one shows that entropy on screen can still be additive if the relation between screen entropy and temperature is given by (\ref{eqn:screen-temperature}).  Some comments are in order:

Firstly, the Parikh-Wilczek is a semiclassical model which could receive further quantum correction.  For instance, a one-loop correction to the surface gravity was considered in the \cite{Banerjee:2008ry}.  As a result, the Bekenstein-Hawking area law receives a logarithmic correction, denoting as $S_{\alpha}(M)=4\pi M^2 -4\pi \alpha \ln{(1+\frac{M^2}{\alpha})}$ and the Hawking temperature is modified as $T_{H}=\frac{M^2+\alpha}{8\pi M^3}$ for large black hole, where coefficient $\alpha$ is related to the trace anomaly.  One can show that the relation of generalized entropic force (\ref{eqn:work-thermal-relation}) is no longer satisfied since 
\begin{equation}
\Delta (T_HS_\alpha) = -\frac{\omega}{2} + \alpha \big[  \frac{3}{2M^2}-\frac{3\alpha+M^2}{2M^4}\ln{(1+\frac{M^2}{\alpha})} \big]\omega.
\end{equation}
Nevertheless, this might be fixed from two different viewpoints.  One way is to modify the Newton's gravitational force by quantum correction up to ${\cal O}(\alpha^2)$, such that $\int{F_{\alpha}(r) dr}$ agrees with the above result.  The other way is to rederive the Hawking temperature, instead of that obtained in the small $\omega$ limit.  The proper definition of Hawking temperature $T_{\alpha}(M)$ is given by 
\begin{equation}
T_{\alpha}(M-\omega)S_{\alpha}(M-\omega)-T_{\alpha}(M)S_{\alpha}(M)=-\frac{\omega}{2}.
\end{equation}
It seems unlikely to have analytic forms for loop corrected $F_{\alpha}(r)$ and $T_{\alpha}(M)$, but at least in principle one can get the order expansion for small $\alpha$.   Once the relation (\ref{eqn:work-thermal-relation}) can be fixed, our conclusion for the temperature-varying screen will still be valid, at least perturbatively in terms of $\alpha$.

Secondly, although Verlinde's adoption of entropic force was meant to tame the wild gravity beast, it is natural to ask how to incorporate other forces in his entropic formalism.   This generalization has been considered for the electromagnetic force, by inclusion of chemical potential in the first law of thermodynamics\cite{Chen:2010ay}.  In particular, for the tunneling model of Reissner-Nordstr\"{o}m black hole radiation \cite{Zhang:2005xt}, one expects the relation (\ref{eqn:work-thermal-relation}) is generalized as
\begin{equation}
F\Delta r = \Delta (TS - AQ).
\end{equation}
where gauge potential and total black hole charges are identified as the chemical potential and number density on the screen respectively.  On the left-hand-side, one has to include electromagnetic force for a consistent result.  If there exists a holographic screen for charged black holes, it must also carry degrees of freedom for charges.  The black hole censorship constraint $Q\le M$ would translate into an upper bound for charge density on the screen.  The break down of screen for excess charge would correspond to possible naked singularity or closed time-like curve.  The recently found charge-mass ratio bound for emission from RN black hole\cite{Kim:2013qxu} seems to support this picture by stating that a part of screen, which corresponds to the radiated particle, also has upper bound for charge density.

\begin{acknowledgments}
WYW is supported in part by the Taiwan's National Science Council under grant NO. 102-2112-M-033-003-MY4 and the National Center for Theoretical Science.
\end{acknowledgments}



\begin{thebibliography}{99}

\bibitem{Verlinde:2010hp} 
  E.~P.~Verlinde,
  ``On the Origin of Gravity and the Laws of Newton,''
  JHEP {\bf 1104}, 029 (2011)

\bibitem{Kobakhidze:2010mn} 
  A.~Kobakhidze,
  ``Gravity is not an entropic force,''
  Phys.\ Rev.\ D {\bf 83}, 021502 (2011)


\bibitem{Chaichian:2011xc} 
  M.~Chaichian, M.~Oksanen and A.~Tureanu,
  ``On gravity as an entropic force,''
  Phys.\ Lett.\ B {\bf 702}, 419 (2011)

\bibitem{Abreu:2013rxe} 
  E.~M.~C.~Abreu and J.~A.~Neto,
  Phys.\ Lett.\ B {\bf 727}, 524 (2013)


\bibitem{Parikh:1999mf}
  M.~K.~Parikh and F.~Wilczek,
  ``Hawking radiation as tunneling,''
  Phys.\ Rev.\ Lett.\  {\bf 85}, 5042 (2000)

\bibitem{Zhang:2009jn}
  B.~Zhang, Q.~-y.~Cai, L.~You and M.~-s.~Zhan,
  ``Hidden Messenger Revealed in Hawking Radiation: A Resolution to the Paradox of Black Hole Information Loss,''
  Phys.\ Lett.\ B {\bf 675}, 98 (2009)

\bibitem{Padmanabhan:2009kr} 
  T.~Padmanabhan,
  ``Equipartition of energy in the horizon degrees of freedom and the emergence of gravity,''
  Mod.\ Phys.\ Lett.\ A {\bf 25}, 1129 (2010)

\bibitem{Tsallis:1988}
C.~Tsallis, 
``Possible generalization of Boltzmann-Gibbs staticis,''
J.\ Stat.\ Phys. {\bf 52} (1988), 1-2, 479-487


\bibitem{Banerjee:2008ry} 
  R.~Banerjee and B.~R.~Majhi,
  ``Quantum Tunneling and Back Reaction,''
  Phys.\ Lett.\ B {\bf 662}, 62 (2008)



\bibitem{Chen:2010ay}
  Y.~-X.~Chen and J.~-L.~Li,
  ``First law of thermodynamics on holographic screens in entropic force frame,''
  Phys.\ Lett.\ B {\bf 700} (2011) 380

\bibitem{Zhang:2005xt}
  J.~-Y.~Zhang and Z.~Zhao,
  ``Hawking radiation of charged particles via tunneling from the Reissner-Nordstroem black hole,''
  JHEP {\bf 0510}, 055 (2005).


\bibitem{Kim:2013qxu} 
  K.~K.~Kim and W.~-Y.~Wen,
  ``Charge-mass ratio bound and optimization in the Parikh-Wilczek tunneling model of Hawking radiation,''
  Phys.\ Lett.\ B {\bf 731}, 307 (2014)

\end{thebibliography}
\end{document}